\documentclass[aps,prl,preprint,tightenlines,superscriptaddress]{revtex4}

\usepackage{graphicx} 
\usepackage{dcolumn}  

\graphicspath{{ps}}

\renewcommand{\arraystretch}{1.1}

\newcommand{\Fig}[1]{Figure~\ref{#1}}

\newcommand{\Tab}[1]{Table~\ref{#1}}

\begin{document}

 \title{ \quad\\[0.5cm]  Extraction of the $b$-quark  shape
 function parameters using the Belle $B \rightarrow X_s  \gamma$
 photon energy spectrum}

\affiliation{University of Melbourne, Victoria}
\affiliation{High Energy Accelerator Research Organization (KEK),
  Tsukuba}

  \author{Antonio~Limosani}\affiliation{University of Melbourne, Victoria} 
  \author{Tadao~Nozaki}\affiliation{High Energy Accelerator Research Organization (KEK), Tsukuba} 
\collaboration{of the Belle Collaboration for the Heavy Flavor
  Averaging Group}

\noaffiliation

\begin{abstract}
We determine the  $b$-quark shape function
parameters, $\Lambda^{\mathrm{SF}}$ and  $\lambda_1^{\mathrm{SF}}$
using the Belle $B \rightarrow X_s  \gamma$ photon energy spectrum.
We assume three models
for the form of the shape function; Exponential, Gaussian and Roman.
\end{abstract}

\maketitle

\section{Introduction}
\label{intro}

The off-diagonal element $V_{ub}$ in the CKM matrix is extracted from measurements of the $B \rightarrow X_u \ell \nu$ 
process in the limited  region of lepton momentum~\cite{CLEOendpt},
or the hadronic recoil mass $M_X$~\cite{BaBar}, or $M_X$ and the
leponic invariant mass squared $q^2$~\cite{Kakuno}  where the contribution of background from the $B \to X_c \ell \nu$ process is suppressed. 
In order to determine $|V_{ub}|$ we need to extrapolate measured rates from such limited regions to the whole phase space. This extrapolation factor is evaluated using a theoretical prediction that takes into account the residual motion of the $b$-quark inside the $B$ meson, so called  ``Fermi motion''~\cite{FN}.  Fermi motion is included in the heavy quark expansion by resumming an infinite set of leading-twist corrections into a shape function of the $b$-quark\cite{Neub,Ural,NeuMan}. Since the shape function is not calculable theoretically, it has to be determined experimentally.

The best way is to make use of the photon energy spectrum for $B \rightarrow X_s  \gamma$  since
both the inclusive decay spectra in  $B \rightarrow X_u \ell \nu$
and  $B \rightarrow X_s \gamma$ are expressed by the same shape
function up to leading order of $1/m_b$ in the heavy quark expansion~\cite{KN,Bigi}. The first results were obtained by CLEO~\cite{Gibbons}, but the errors of the shape function parameters are rather large.  Therefore the uncertainty of the shape function dominates the theoretical error of $|V_{ub}|$ at present. Belle has recently provided more precise data than CLEO of the $B
\rightarrow X_s \gamma$ photon  spectrum~\cite{Koppenburg}.
We report on the results of determination of the shape function parameters using the Belle $B \rightarrow X_s \gamma$ data.

\section{Procedure}
We used a method based on that devised by the CLEO
collaboration\cite{Anderson}. We fit Monte Carlo (MC)  simulated spectra to the raw data
photon energy spectrum. ``Raw'' refers to the spectra that are obtained
after the application of the $B\rightarrow X_s \gamma$ analysis
cuts. The use of ``raw'' spectra correctly accounts for the
Lorentz boost from the $B$ rest frame to the center of mass system,  energy resolution effects and avoids unfolding. The
method is as follows;
\begin{enumerate}
\item Assume a shape function model.
\item Simulate the photon energy spectrum for a certain set of
  parameters; $(\Lambda^{\mathrm{SF}},\lambda_1^{\mathrm{SF}})$.
\item Perform a $\chi^2$ fit of the  simulated spectrum to the data
where only the normalization of the simulated spectrum is floated and keep the resultant $\chi^2$ value.
\item Repeat steps 2-3 for different sets of parameters to construct a two dimensional grid with each point having a $\chi^2$.  
\item Find the minimum  $\chi^2$ on the grid and all points on the grid that are one unit of  $\chi^2$ above the minimum.
\item Repeat steps 1-5 for a different shape function model.
\end{enumerate}

\subsection{Shape function models}
\begin{figure}[hbpt!]
    \begin{center}
      \begin{tabular}{cc}
      \includegraphics[width=0.50\columnwidth]{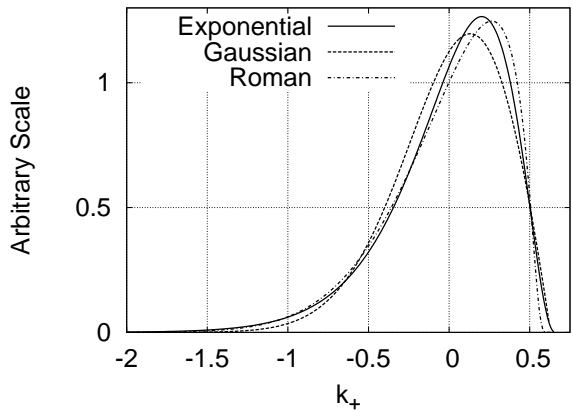} 
      \end{tabular}
    \end{center}
    \caption[Shape function forms]{Shape function model curves for
      Exponential
      $(\Lambda^{\mathrm{SF}},\lambda_1^{\mathrm{SF}})=(0.66,-0.40)$,
      Gaussian
      $(\Lambda^{\mathrm{SF}},\lambda_1^{\mathrm{SF}})=(0.63,-0.40)$,
      and 
      Roman
      $(\Lambda^{\mathrm{SF}},\lambda_1^{\mathrm{SF}})=(0.66,-0.39)$,
      where  $\Lambda^{\mathrm{SF}}$ and $\lambda_1^{\mathrm{SF}}$
      are measured in units of $\mathrm{GeV}/c^2$ and
      $\mathrm{GeV}^2/c^2$ respectively.\label{fig:sfcurves}}
  \end{figure}

Three shape function forms suggested in the literature are employed;
Exponential, Gaussian and Roman\cite{Bigi,KN}. These are
described in \Tab{tab:sfforms}. The  shape function $F$ is a function of 
$k_+(\equiv k_0 - k_3)$, where $k_\mu$ is the residual momentum of
the $b$-quark in the $B$ meson,
defined through
\begin{equation}
  p_{b,\mu}=m_bv_\mu + k_\mu,
\end{equation} where $v_\mu = (1,0,0,0)$ and $k_3$
is the $k$ component along the direction of the $u$-quark.
The shape function is parameterized by $\Lambda^{\mathrm{SF}}$ and
$ \lambda_1^{\mathrm{SF}}$. These parameters
are related to the $b$ quark mass, $m_b$, and
the average momentum squared of the $b$ quark, $\mu_\pi^2$, via the
relations,
\begin{equation}
  \Lambda^{\mathrm{SF}} = M_B -m_b
\end{equation}  
and
\begin{equation}
  \lambda_1^{\mathrm{SF}} = -\mu_\pi^2,
\end{equation}
where $M_B$ is the mass of the $B$ meson. Up to leading order in the
non-perturbative dynamics the shape function is universal in
describing the $b$-quark Fermi motion relevant to $b$-to-light quark
transitions. The lepton and photon energy spectra in $B\rightarrow
X_u l \nu$ and $B\rightarrow X_s\gamma$ decays are given by the
convolution of the respective parton-level spectra with the shape
function. 
Example shape function curves are plotted in \Fig{fig:sfcurves}(a). 

\begin{table}[hbpt!]
  \begin{center}{
      \renewcommand{\arraystretch}{1.8}
      \begin{tabular}{|c|c|} \hline
        Shape Function & Form \\ \hline
        Exponential
        & $F(k_+;a) = N (1-x)^a e^{(1+a)x} $ \\  \hline
        Gaussian
        & $F(k_+;c) = N (1-x)^c e^{-b(1-x)^2} $ \\
        & where $b=\left(\Gamma(\frac{c+2}{2})/\Gamma(\frac{c+1}{2})\right)^2$ \\ \hline
        Roman
        & $F(k_+;\rho) = N \frac{\kappa}{\sqrt{\pi}} \exp \big \{
        -\frac{1}{4} \big ( \frac{1}{\kappa} \frac{\rho}{1-x} -
        \kappa (1-x) \big )^2  \big \}$ \\
        
        & where $\kappa=\frac{\rho}{\sqrt{\pi}}
        e^{\rho/2}K_1(\rho/2)$ \\ \hline
        
        \multicolumn{2}{|c|}{where $x=k_+ / \Lambda^{\mathrm{SF}}$} \\
        \multicolumn{2}{|c|}{$-m_b \leq k_+ \leq \Lambda^{\mathrm{SF}}$} \\ 
        \multicolumn{2}{|c|}{and $a,c,\rho, N$ are chosen} \\
        \multicolumn{2}{|c|}{to satisfy} \\
        \multicolumn{2}{|c|}{$A_0=1$, $A_1=0$,
        $A_2=-\lambda_1^{\mathrm{SF}}/3$, } \\
        \multicolumn{2}{|c|}{where $A_n=\int k_+^n F(k_+) dk_+$} \\ \hline
      \end{tabular}
      }
    \end{center}
    \caption[Shape function models]{The three models used for the shape function forms\label{tab:sfforms}}
  \end{table}

  \subsection{Monte Carlo simulated photon energy spectrum}
We generate $B\rightarrow X_s \gamma$ MC events according to  
 the Kagan and Neubert prescription for each set of the shape function parameter values~\cite{KN}. The generated events are then simulated for the detector performance using the Belle detector simulation program.
Afterwards $B\rightarrow X_s \gamma$ analysis cuts are applied to the MC events to obtain the raw photon energy spectrum in the $\Upsilon(4S)$ rest frame~\cite{Koppenburg}.

   \subsection{Fitting the spectrum}
\begin{figure}
     \begin{center}
       \includegraphics[width=0.50\columnwidth]{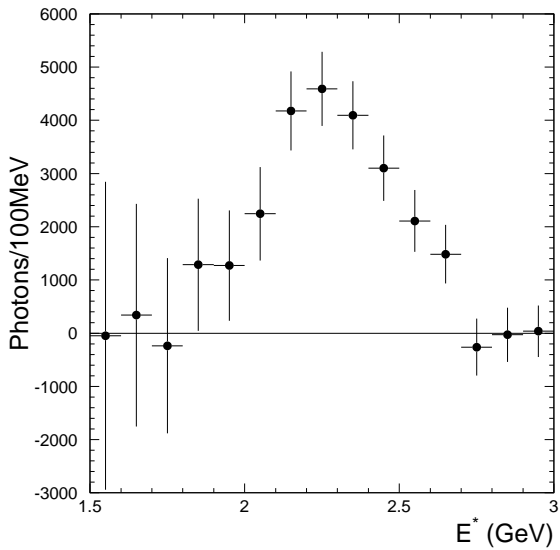} \\
     \end{center}
     \caption{Raw $B \rightarrow X_s \gamma$ photon energy spectra in
       the $\Upsilon(4S)$ frame as acquired from data. The
       errors include both statistical and systematic errors.
       Raw refers to spectra as measured after the application of
       Belle  $B \rightarrow X_s \gamma$ analysis cuts.
       \label{fig:bsgraw}}
   \end{figure}

   For a given set of shape function parameters,
   a $\chi^2$ fit of the MC simulated photon spectrum to the raw data spectrum
   is performed in the interval, $1.5 <E^*_\gamma/\mathrm{GeV} <
   2.8$\footnote{The $*$ denotes the center of mass frame or
   equivalently the $\Upsilon(4S)$ rest frame}.
   The normalization parameter is floated in the fit. The
   raw spectrum is plotted in \Fig{fig:bsgraw}, the errors include both statistical and
   systematic errors. The latter are dominated by the
   estimation of the $B\bar{B}$ background and are 100\% correlated.
   Therefore the covariance matrix is constructed as 
   \begin{equation}
     V_{ij} =
     \sigma^{\mathrm{stat}}_{d_i}\sigma^{\mathrm{stat}}_{d_j}\delta_{ij}
     +   \sigma^{\mathrm{sys}}_{d_i}\sigma^{\mathrm{sys}}_{d_j}  
   \end{equation}
   where $i,j=1,2,\ldots,13$ denote the bin number, and
   $\sigma_d$ is the error in the data. 
   Then the $\chi^2$ used in the fitting is given by
   \begin{equation}
     \chi^2 = \sum_{ij} (d_i-f_i) (V_{ij})^{-1}  (d_j-f_j), 
   \end{equation}
   where $(V_{ij})^{-1}$ denotes the $ij^{th}$ element of the inverted
   covariance matrix.
   The $\chi^2$ value after the fit is used to determine a map of $\chi^2$ 
as a function of the shape function  parameters.

   \subsection{The best fit and $\Delta \chi^2$ contour}
   The best fit parameters are associated to the minimum
   chi-squared case, $\chi^2_\mathrm{min}$. The $1\sigma$ ``ellipse'' is
   defined as the contour which satisfies $\Delta\chi^2
   \equiv \chi^2-\chi^2_\mathrm{min}=1$.
   The contours are found to be well approximated by the function\cite{Fac}, 
   \begin{equation}\label{eqn:ellipse}
     \Delta \chi^2(\Lambda^{\mathrm{SF}},\lambda_1^{\mathrm{SF}}) = \left ( \frac{\lambda_1^{\mathrm{SF}} + a
         (\Lambda^{\mathrm{SF}})^2 + b }{c} \right )^2 +
     \left ( \frac{(\Lambda^{\mathrm{SF}})^2 + d}{e} \right )^2. 
   \end{equation}
   The parameters $a$, $b$, $c$, $d$, and $e$ are determined by fitting the
   function to the parameter points that lie on the contour. 

   \section{Results}
\begin{figure}[hbpt!]
\begin{center}
  \begin{tabular}{ccc}
    \includegraphics[width=0.33\columnwidth]{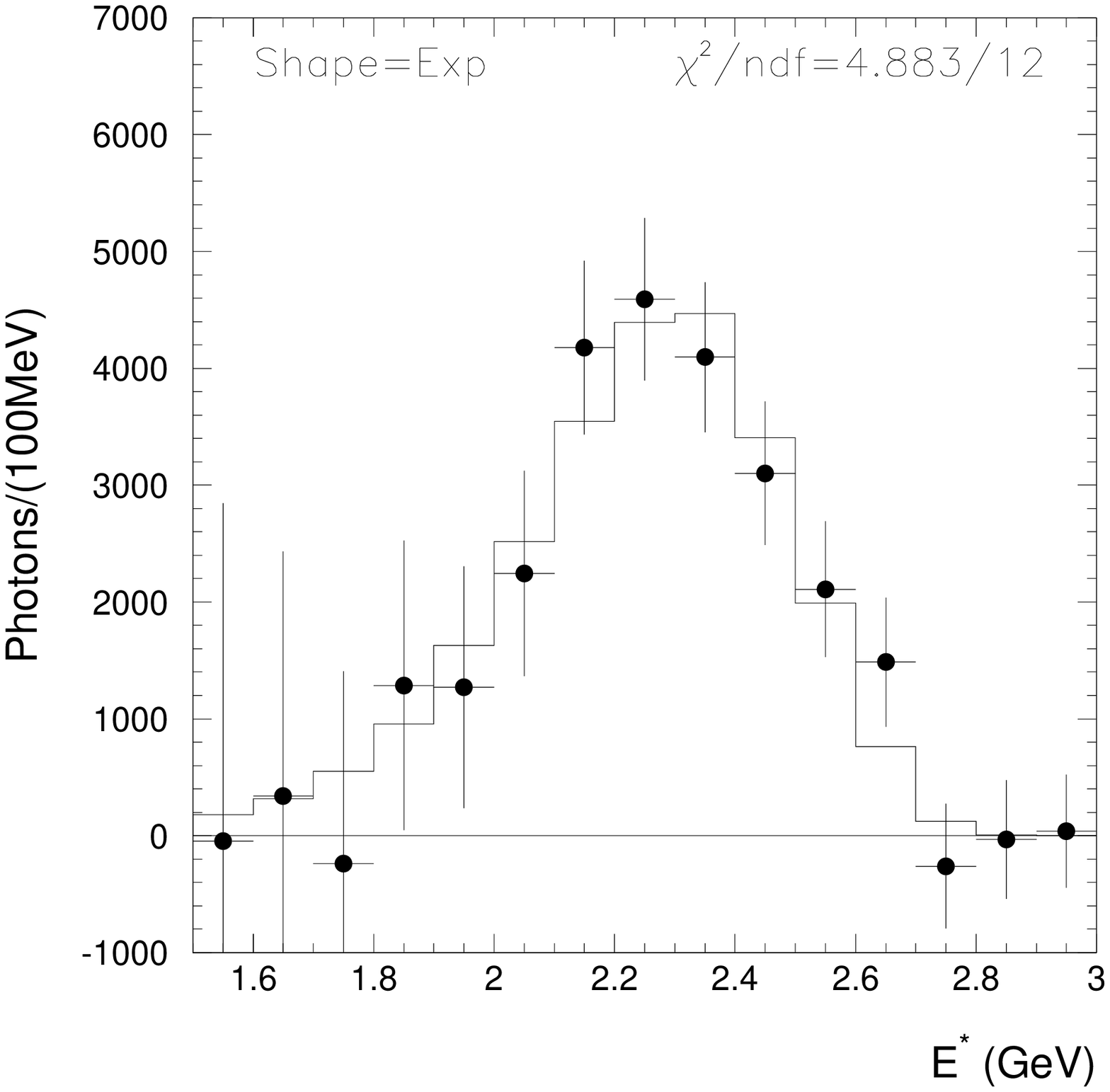} & 
    \includegraphics[width=0.33\columnwidth]{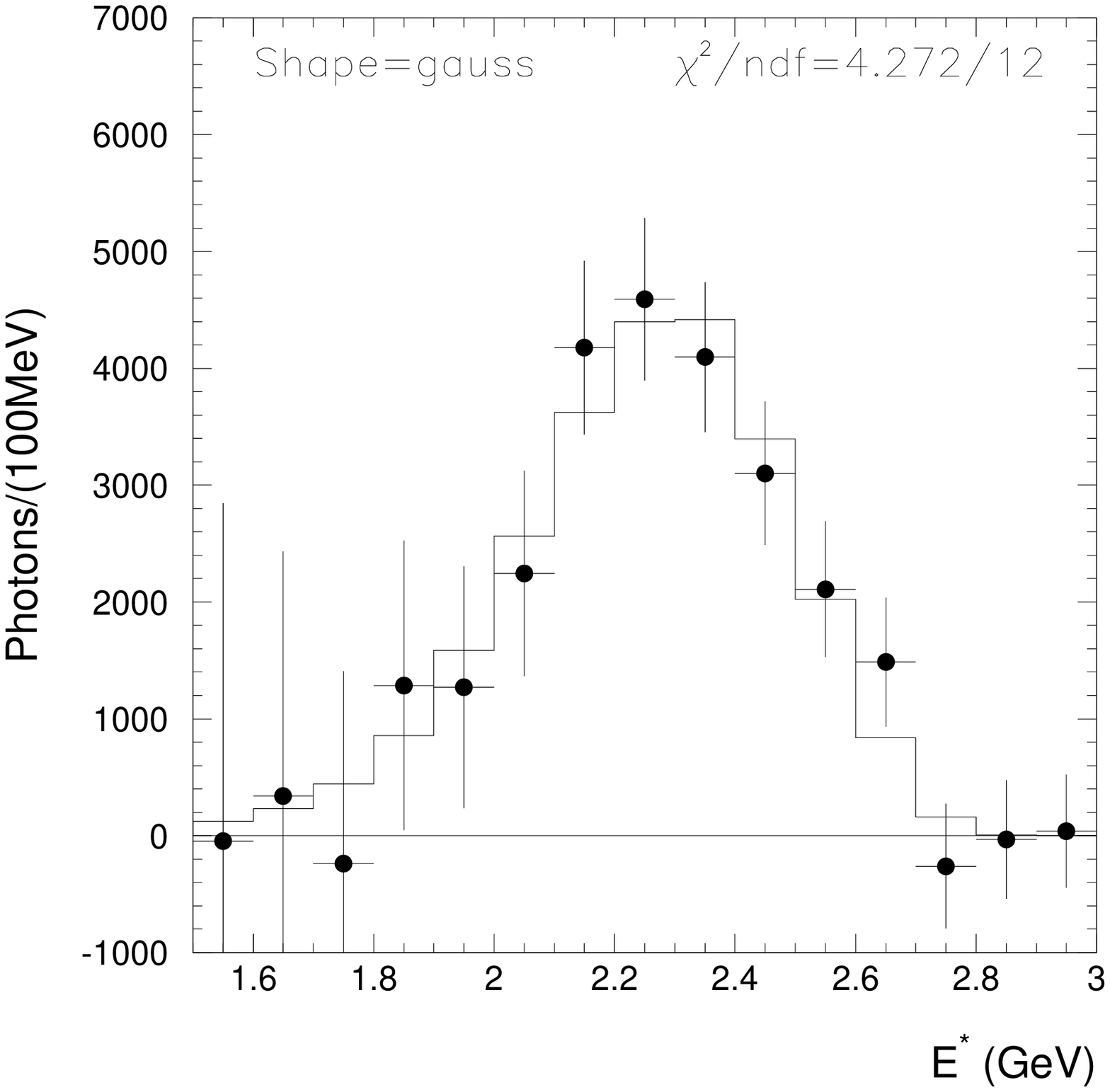} &
    \includegraphics[width=0.33\columnwidth]{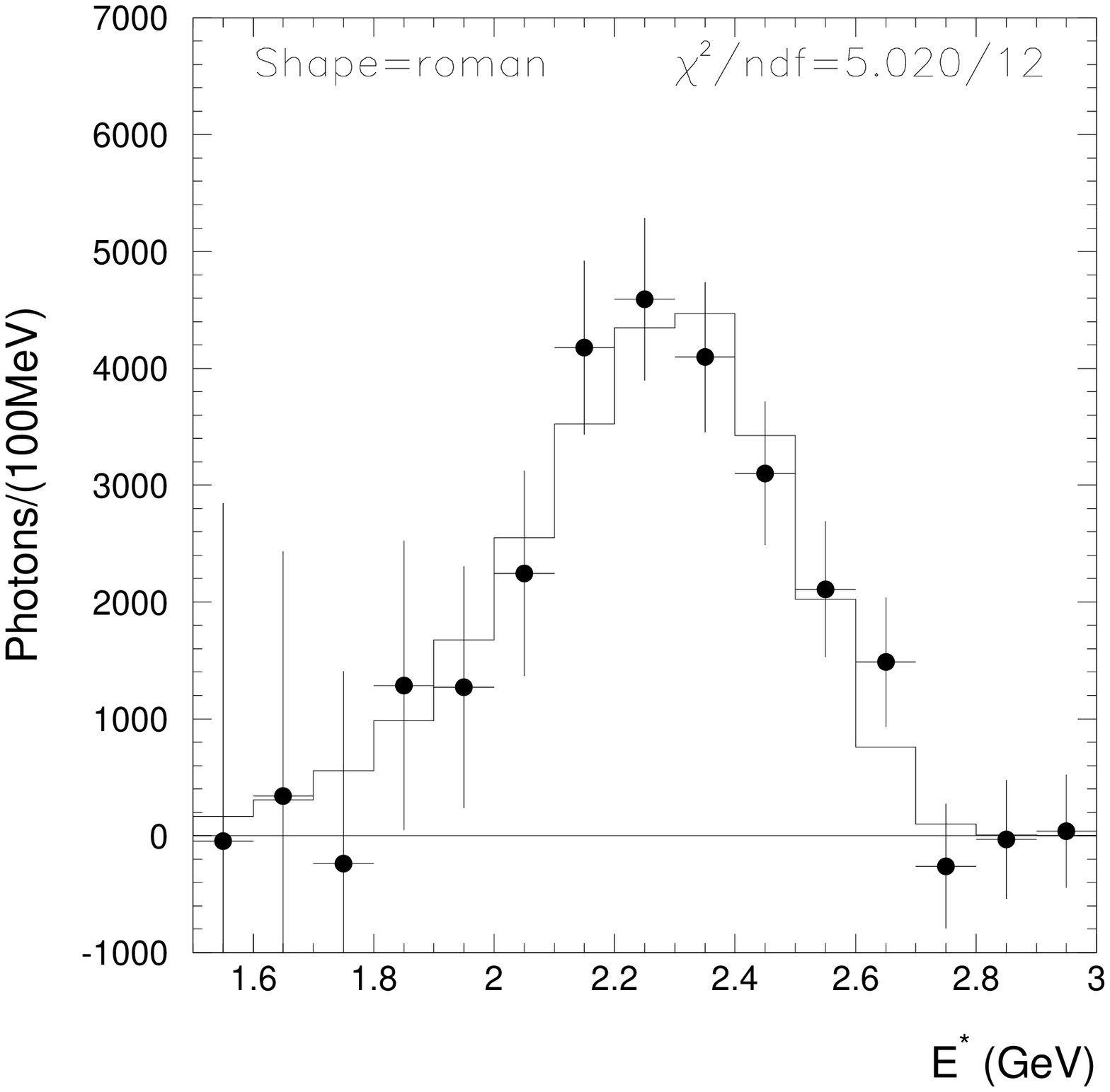} \\
    (a) Exponential & (b) Gaussian & (c) Roman \\
  \end{tabular}
\end{center}
\caption[The minimum $\chi^2$ fits]{The  minimum $\chi^2$ fits of MC
    simulated spectra to the raw data for each shape function model
    \label{fig:bestfit}}
\end{figure}

\begin{figure}[hbpt!]
  \begin{center}
    \begin{tabular}{ccc}
      \includegraphics[width=0.33\columnwidth]{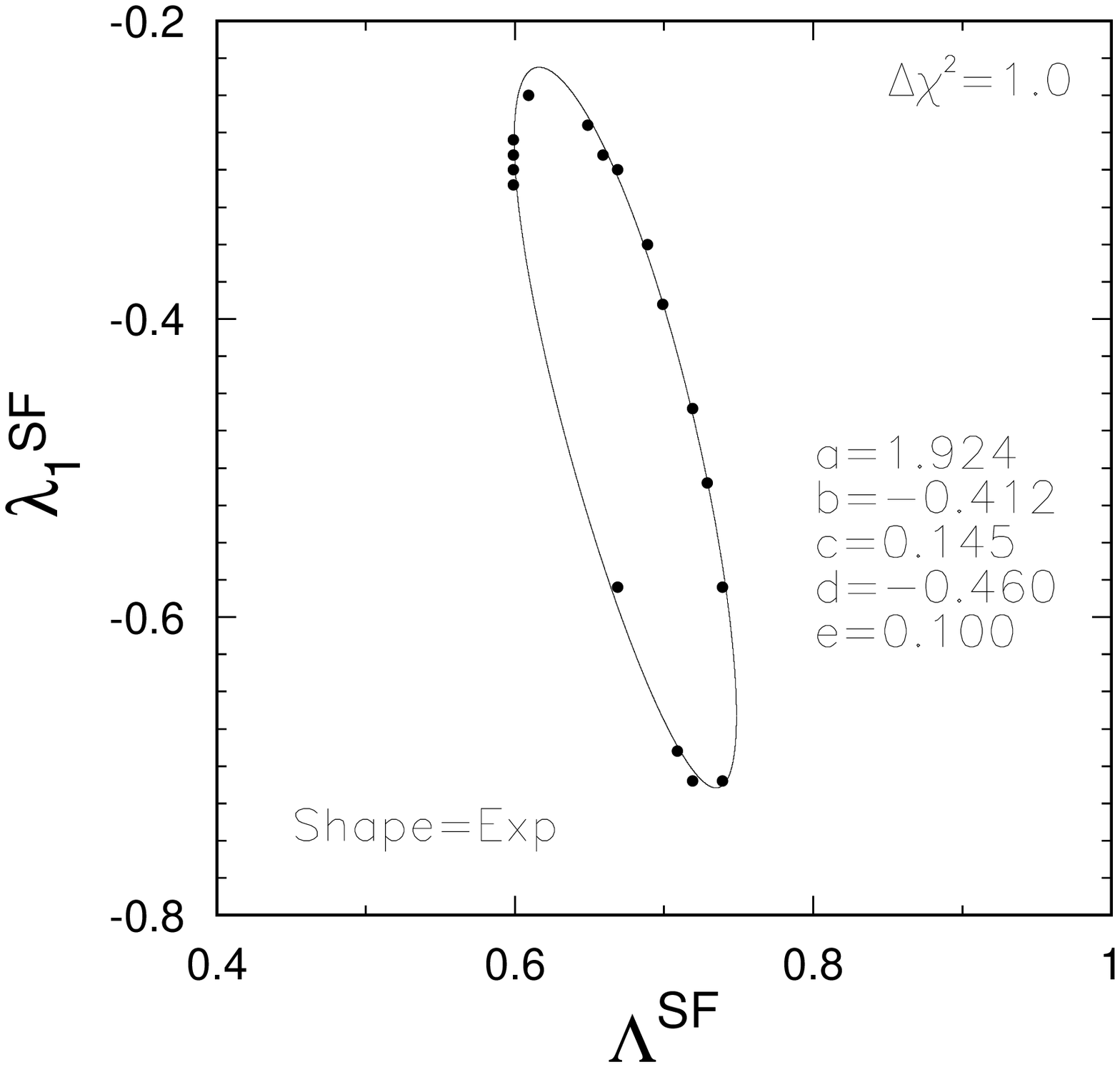} & 
      \includegraphics[width=0.33\columnwidth]{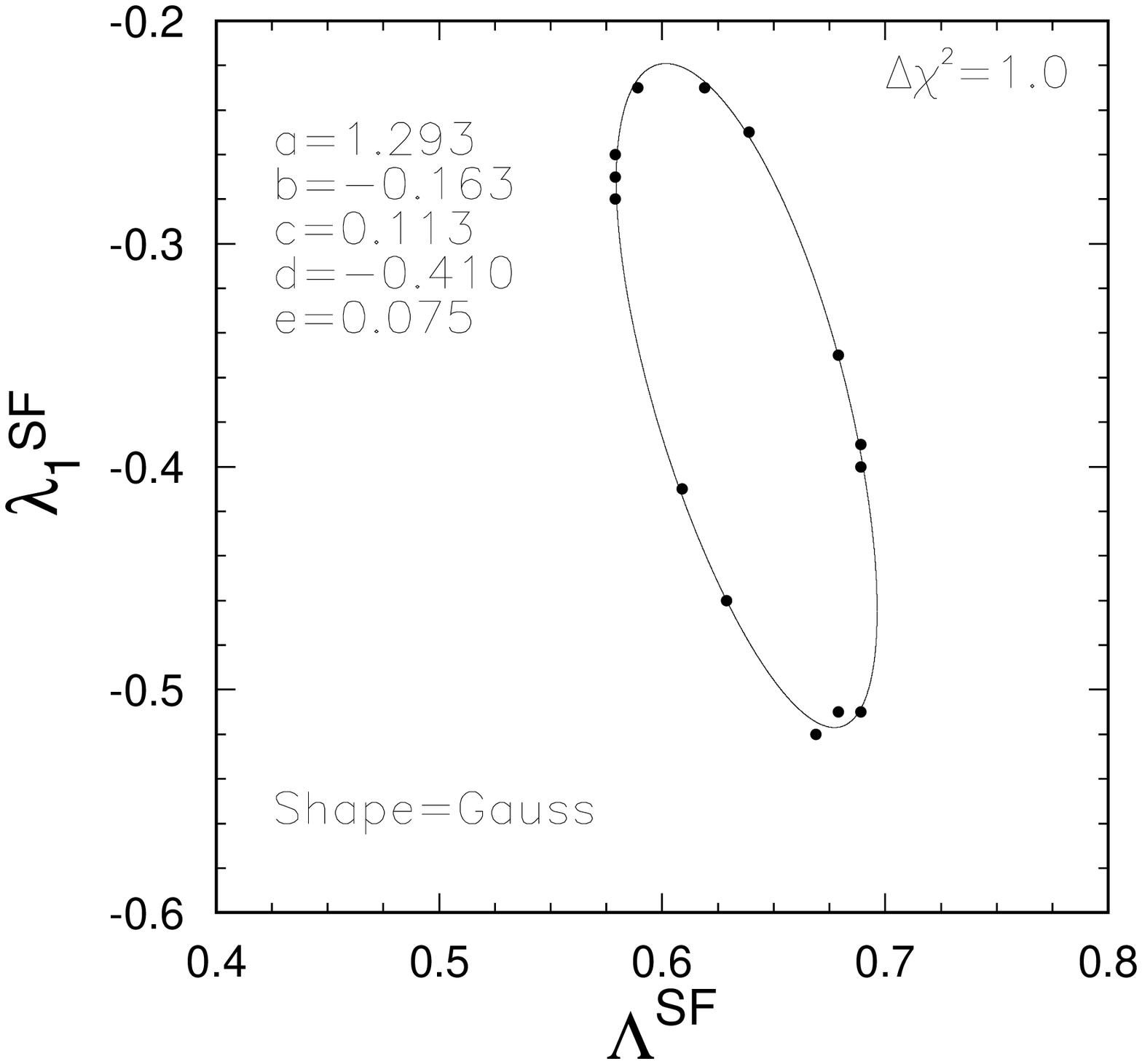} & 
      \includegraphics[width=0.33\columnwidth]{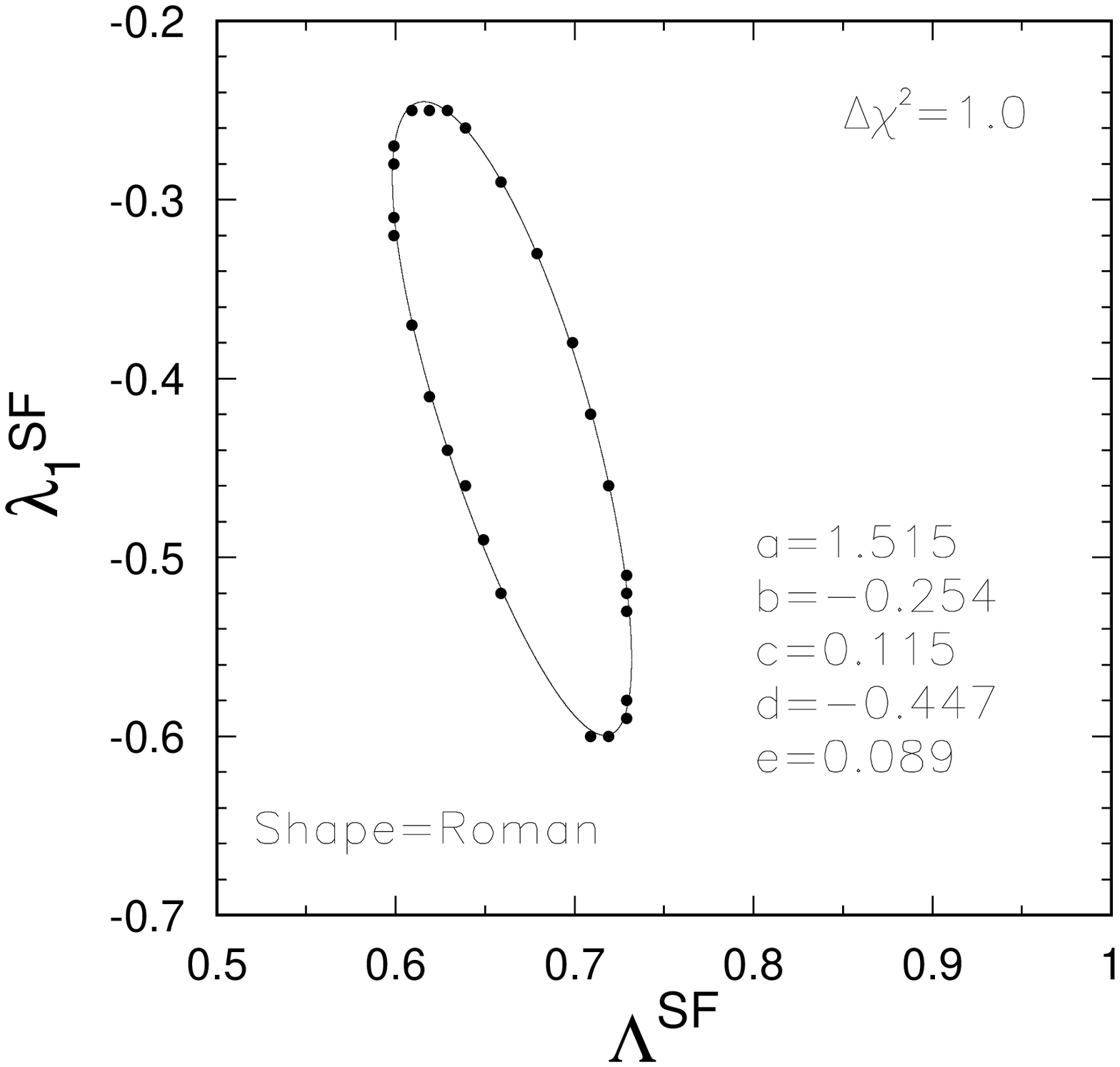}
      \\
      (a) Exponential & (b) Gaussian & (c) Roman \\
    \end{tabular}
  \end{center}
  \caption[The fitted $\Delta\chi^2=1$ contours]{The fitted
      $\Delta\chi^2=1$ contours for each shape function model \label{fig:contour}}
\end{figure}

   The best fit parameters are given in
   \Tab{tab:bestfit}. The parameter values are found to be consistent
   across all three shape function forms. The minimum $\chi^2$ fit for
   each shape function model is displayed in
   \Fig{fig:bestfit}. The fits to the contour with
   $\Delta\chi^2=1$ points are shown in \Fig{fig:contour}. The imposed
   shape function form acts to correlate $\Lambda^{\mathrm{SF}}$ and
   $\lambda_1^{\mathrm{SF}}$. 

\begin{table}[hbpt]
  \begin{center}
  \begin{tabular}{|l|c|c|c|} \hline                      
    Shape   &$\chi^2_\mathrm{min}$ & $\Lambda^{\mathrm{SF}}$ & $\lambda_1^{\mathrm{SF}}$  \\ 
     &  & $(\mathrm{GeV}/c^2)$ & $(\mathrm{GeV}^2/c^2)$  \\ \hline
    Exponential & 4.883 & 0.66  & -0.40  \\
    Gaussian    & 4.272 & 0.63  & -0.33  \\
    Roman       & 5.020 & 0.66  & -0.39  \\ \hline
  \end{tabular} 
  \end{center}
  \caption{The best fit shape function parameter values\label{tab:bestfit}}
\end{table}

\subsection{Strong Coupling $\alpha_s$}
The strong coupling constant, $\alpha_s$, is an input into the
parton-level calculations for both $B\rightarrow X_s \gamma$ and
$B \rightarrow X_u l \nu$ spectra. By default $\alpha_s(\mu)$ is
evaluated at the mass scale $\mu=m_b$.
To investigate the systematic effect of this
choice the analysis is redone for $\mu=m_b/2$ and $\mu=2m_b$
in the case of the exponential
shape function model. The $\Lambda^{\mathrm{SF}}$ and
$\lambda_1^{\mathrm{SF}}$ parameter values corresponding to 
$\chi^2_{\mathrm{min}}$ are given in \Tab{tab:alphas}.

\begin{table}[hbpt]
  \begin{center}
    \begin{tabular}{|l|c|c|c|} \hline                      
      $\mu$   &$\alpha_s(\mu)$ & $\Lambda^{\mathrm{SF}}$ & $\lambda_1^{\mathrm{SF}}$  \\ 
         &  & $(\mathrm{GeV}/c^2)$ & $(\mathrm{GeV}^2/c^2)$   \\ \hline
      $m_b$   & 0.210 & 0.66  & -0.40  \\
      $m_b/2$ & 0.257 & 0.65  & -0.41  \\
      $2m_b$  & 0.177 & 0.68  & -0.43  \\ \hline
    \end{tabular} 
  \end{center}
  \caption{The best fit parameters for various $\alpha_s$ using the
      exponential shape function model\label{tab:alphas}}
\end{table}

\section{Comparison with CLEO}
The CLEO collaboration has provided points which lie on their
equivalent $\Delta\chi^2=1$ contour for the case of an exponential
shape function model\cite{Hen}.
The data points are slightly different from those given in the Gibbons' report\cite{Gibbons} since the present data now includes the uncertainty in the
$B\bar{B}$ background Monte Carlo normalization\cite{Hen}.

\begin{figure}[hbpt!]
  \begin{center}
      \includegraphics[width=0.50\columnwidth]{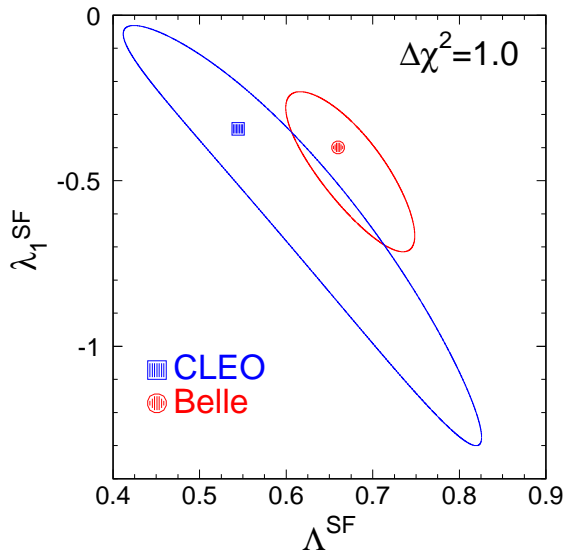} 
  \end{center}
  \caption{The fitted
      $\Delta\chi^2=1$ contours for CLEO (blue) and Belle (red)
      assuming an exponential shape function form.\label{fig:comb}}
\end{figure}

We fit the functional form given in
equation~\ref{eqn:ellipse} to their contour data and find excellent
agreement ($a=2.378$, $b=-0.347$, $c=0.178$, $d=-0.426$, $e=0.256$).
The minimum $\chi^2$ point for the CLEO data corresponds to
$(\Lambda^{\mathrm{SF}},\lambda_1^{\mathrm{SF}})_{\mathrm{Exp}}=(0.545,-0.342)$.
We compare  the CLEO and Belle contours in \Fig{fig:comb}. 
The regions bounded by the contours marginally
overlap. The uncertainty in the Belle result is much reduced with
respect to that of CLEO. 

Unfortunately we can not produce a combined $\Delta\chi^2=1$ contour
of the two experiments since a precise map of $\Delta\chi^2$ as a function
of $\Lambda^{\mathrm{SF}}$ and $\lambda_1^{\mathrm{SF}}$ 
is not currently available for CLEO.

\section{Summary}
We have determined the $b$-quark shape function
parameters, $\Lambda^{\mathrm{SF}}$ and  $\lambda_1^{\mathrm{SF}}$,
from fits of Monte Carlo simulated spectra to the raw Belle measured $B\rightarrow
X_s \gamma$ photon energy spectrum. Raw refers to the spectrum as
measured after the application of analysis cuts. We used three models
for the form of the shape function; Exponential, Gaussian and Roman.
We found the best fit parameters;
$(\Lambda^{\mathrm{SF}},\lambda_1^{\mathrm{SF}})_{\mathrm{Exp}}=(0.66,-0.40)$,
$(\Lambda^{\mathrm{SF}},\lambda_1^{\mathrm{SF}})_{\mathrm{Gauss}}=(0.63,-0.33)$,
and   
$(\Lambda^{\mathrm{SF}},\lambda_1^{\mathrm{SF}})_{\mathrm{Roman}}=(0.66,-0.39)$,
where $\Lambda^{\mathrm{SF}}$ and $\lambda_1^{\mathrm{SF}}$
are measured in units of $\mathrm{GeV}/c^2$ and $\mathrm{GeV}^2/c^2$ respectively.
We also determined the $\Delta \chi^2 = 1$ contours in the
$(\Lambda^{\mathrm{SF}},\lambda_1^{\mathrm{SF}})$ parameter space for
each of the assumed models.

\section*{ACKNOWLEDGEMENTS}
We would like to thank all Belle collaborators, in particular Patrick Koppenburg. 
We acknowledge support from the Ministry of Education,
Culture, Sports, Science, and Technology of Japan;
the Australian Research Council
and the Australian Department of Education, Science and Training.


\begin{thebibliography}{99}
\bibitem{CLEOendpt}
  A. Bornheim \textit{et al.} (CLEO Collaboration), Phys. Rev. Lett. \textbf{88}, 231803 (2002).
\bibitem{BaBar}
  B. Aubert \textit{et al.} (BaBar Collaboration), Phys. Rev. Lett. \textbf{92}, 071802 (2004).
\bibitem{Kakuno}
  H. Kakuno \textit{et al.} (Belle Collaboration), Phys. Rev. Lett. \textbf{92}, 101801 (2004).
\bibitem{FN}
   F.D. Fazio and  M. Neubert, J. High Energy Phys. \textbf{9906}, 017 (1999).
\bibitem{Neub}
   M. Neubert, Phys. Rev. \textbf{D 49}, 3392 and 4623 (1994).
\bibitem{Ural}
   I. Bigi, M. Shifman, N. Uraltsev and A. Vainshtein, Int. J. Mod. Phys. \textbf{A 9}, 2467 (1994). R. Dickman, M. Shifman, and N. Uraltsev, Int. J. Mod. Phys. \textbf{A 11}, 571 (1996).
\bibitem{NeuMan}
   T. Mannel and M. Neubert, Phys. Rev. \textbf{D 50}, 2037 (1994).
\bibitem{Bigi}
   I. Bigi, M. Shifman, N. Uraltsev and A. Vainstein, Phys. Lett. \textbf{B328},  431 (1994).
\bibitem{KN}
   A.L.  Kagan and M. Neubert, Eur. Phys. J. \textbf{C7} 5 (1999). 
\bibitem{Gibbons}  
L. Gibbons (CLEO Collaboration),  hep-ex/0402009.
\bibitem{Koppenburg}
  P. Koppenburg  \textit{et al.} (Belle Collaboration), hep-ex/0403004.
\bibitem{Anderson}  
 S. Anderson, Ph.D. thesis, University of Minnesota, 2002.
\bibitem{Fac}  
 We thank R. Faccini for suggesting such a function.
\bibitem{Hen}  
 D. Cronin-Hennessy, private communication.
\end{thebibliography}
\end{document}